\documentclass[twocolumn,aps,prl,10pt,floatfix,superscriptaddress]{revtex4-2}
\usepackage{graphicx}
\usepackage[usenames,dvipsnames]{xcolor}
\usepackage[colorlinks=true,citecolor=blue,linkcolor=blue,urlcolor=blue,filecolor=blue,linktocpage=true]{hyperref}
\usepackage{textcomp}
\usepackage{stackengine}
\usepackage{listings}
\usepackage[makeroom]{cancel}
\usepackage[normalem]{ulem}        
\usepackage{array}
\usepackage{enumerate}             
\usepackage[shortlabels]{enumitem}  
\usepackage{verbatim}

\usepackage{amsmath}
\usepackage{amsfonts}
\usepackage{amssymb}
\usepackage{bbold}
\usepackage{relsize}
\usepackage{mathrsfs}
\usepackage{mathtools}
\usepackage{IEEEtrantools}          

\usepackage{orcidlink}

\usepackage[utf8x]{inputenc}
\usepackage{bm}
\usepackage{braket}
\usepackage{siunitx}
\usepackage{soul}
\usepackage{dcolumn} 
\usepackage{physics} 
\usepackage{dsfont}  

\usepackage{txfonts}   
\usepackage{xr-hyper}

\usepackage[version=4]{mhchem}      
\usepackage{textcase}               

\begin{document}
\title{
Nucleating a Different Coordination in a Crystal under Pressure: \\a Study of the $B$1-$B$2 Transition in NaCl by Metadynamics}
\author{Matej Badin\,\orcidlink{0000-0001-7487-9802}}
\email{mbadin@sissa.it}
\affiliation{SISSA – Scuola Internazionale Superiore di Studi Avanzati, Via Bonomea 265, 34136 Trieste, Italy}
\affiliation{Department of Experimental Physics, Comenius University, Mlynsk\'{a} Dolina F2, 842 48 Bratislava, Slovakia}

\author{Roman Marto\v{n}\'{a}k\,\orcidlink{0000-0002-4013-9117}}
\email{martonak@fmph.uniba.sk}
\affiliation{Department of Experimental Physics, Comenius University, Mlynsk\'{a} Dolina F2, 842 48 Bratislava, Slovakia}

\date{\today}
\begin{abstract}
Here we propose an NPT metadynamics simulation scheme for pressure-induced structural phase transitions, using coordination number and volume as collective variables, and apply it to the
reconstructive structural transformation $B$1/$B$2 in NaCl. By studying systems with size up to 64 000 atoms we reach a regime beyond collective
mechanism and observe transformations proceeding via nucleation and growth. We also reveal the crossover of the transition mechanism from Buerger-like for smaller systems to Watanabe/Tol\'{e}dano for larger ones.
The scheme is likely to be applicable to a broader class of pressure-induced structural transitions, allowing study complex nucleation effects and
bringing simulations closer to realistic conditions.
\end{abstract}
\maketitle
Structural phase transitions in crystals induced by pressure or temperature are 
complex phenomena of great fundamental and practical importance. 
Most of them are reconstructive, thermodynamically first order, and involve crossing of free-energy barriers via a non-trivial concerted
motion of atoms, representing a rare event. These transitions give rise to a number of important phases with unique properties such as, e.g.,~diamond
created from graphite at high-pressure conditions. In the process of synthesis of such phases, kinetics plays a key role in determining the outcome
of the transition, which might not necessarily be the thermodynamically most stable form, but rather a metastable one (e.g. after compression of 
silicon in the cubic-diamond structure to 11 GPa and decompression to ambient pressure, the BC8 phase is found)~\cite{Wentorf1963}.

In the past two decades spectacular progress has been made in the prediction of crystalline phases, due to the advent of methods such as evolutionary
search~\cite{Oganov-Glass-USPEX}, random search~\cite{PhysRevLett.97.045504,Pickard_2011}, particle swarm optimisation~\cite{Wang2010},
minima hopping~\cite{Goedecker2004}, etc. These approaches very effectively address the thermodynamics of the problem, identifying stable and
metastable structures as global or local minima of the enthalpy surface. However, an understanding the mechanisms of the transitions, the pertinent barriers 
in the free-energy surface (FES) and the resulting kinetics still lags behind and more detailed information about the FES is needed to make progress. 
Commonly used theoretical approaches to uncover possible mechanisms and estimate energetic barrier per unit cell are based on geometric 
modelling~\cite{Shoji1931,smoluchowski,HydeOKeeffe,Watanabe1977}, group-theory~\cite{Pendas1994,Stokes2002,Stokes2004,Sims1998}, phenomenological Landau 
theory~\cite{Toldano2003}, or, more recently, exploration of the FES~\cite{Shang2013,C3CP44063J,ISHIKAWA201436}. However, by assuming collective transformation throughout the crystal, 
they cannot by construction assess the size of the nucleation region and determine the true nucleation barrier. A realistic simulation must therefore reach 
beyond collective behaviour and include nucleation. We note that one of the methods allowing mapping of FES, metadynamics (MetaD)~\cite{Laio2002} (for recent review see Ref.~\cite{Bussi2020}) was successfully
applied to the problem of crystallisation from liquid~\cite{Niu2018,Piaggi2017,Gobbo2018,Piaggi2018,PiaggiEnthalpyPRL2017,Piaggi2019} which has a number of
similar features to the problem of solid-solid transitions. 

The application of MetaD to structural transitions in crystals started in Refs.~\cite{Martonak2003, Martonak2006}, using the h-matrix of
the supercell vectors (similarly to the Parrinello-Rahman variable-cell MD~\cite{PR1, PR2}) as the generic 6D collective variable (CV). This approach is efficient in inducing structural transitions in a number of 
systems~\cite{Martonak2005,Raiteri2005,Martonak2007,Behler2008,Bealing2009,Sun2009,Yao2009,CO2_2009,Hromadova2011,Plasienka2015,Tong2021,csp_wiley2010,epjb2011}, however, the use of a 6D CV essentially limits 
the use of MetaD to escaping FES minima and precludes the FES reconstruction. For an efficient reconstruction of the FES~\cite{Laio2005} CVs with dimensionality up to 3 are usually chosen. 
Moreover, the supercell-based CV by construction works well only for relatively small systems where transitions proceed via collective mechanisms but is 
unlikely to allow the study of nucleation in a large system. Several approaches addressing an autonomous construction
or a choice of CVs have been proposed recently~\cite{Tiwary2016,Mones2016,McCarty2017,MSultan2017,Gimondi2018,Rodriguez2018,Zhang2018,Sultan2018,Mendels2018,Rizzi2019}.  
Applications of MetaD to structural transitions not based on the supercell CV are presented in 
Refs.~\cite{Zipoli2004,Pipolo2017,Gimondi2017,Jobbins2018,MendelsAgI2018,Rogal2019}.

We present in this Letter a simple and general scheme based on physically motivated CVs such as coordination number (CN) and volume (V) that should be applicable to
the important class of pressure-induced structural transitions. This choice is primarily motivated by one of generic rules of high-pressure chemistry formulated by Prewitt and 
Downs~\cite{Hemley1998,Grochala2007,Miao2020} that states that pressure-induced transitions are typically accompanied by an increase of CN in the 
$1^\mathrm{st}$ coordination sphere. In a more general context, CN was proposed as a reaction coordinate in constrained MD in Ref.~\cite{Sprik1998}. It was also employed in an 
early MetaD study of a structural transition in carbon~\cite{Zipoli2004} and in a MetaD study of the $B$1-$B$2 transition in colloidal clusters~\cite{Bochicchio2014}. Thermodynamically, 
in first-order transitions an abrupt densification of the system takes place, with a jump in volume from a few \% up to 10-20\%. We show here that the combination of 
CN \& V appears to provide an effective 2D CV able to drive pressure-induced structural transition.

We demonstrate the applicability of this scheme on the pressure-induced $B$1-$B$2 transition in \ce{NaCl}, which represents a paradigmatic but also very complex example 
of a reconstructive transition~\cite{Toldano2003}. It occurs at room temperature at $p= \SI{26.6}{\giga\pascal}$ and involves a volume drop of 5\%~\cite{Li1987}. 
Several theoretical collective mechanisms were proposed for this transition, falling essentially into two groups. The ones by Shoji~\cite{Shoji1931}, Buerger~\cite{smoluchowski}
and Stokes \& Hatch~\cite{Stokes2002} are mainly driven by lattice strain while the other class by Hyde \& O'Keeffe~\cite{HydeOKeeffe}, Watanabe \textit{et al.} 
(WTM)~\cite{Watanabe1977} and Toledáno \textit{et al.}~\cite{Toldano2003} involves more shuffling of atoms~\cite{Stokes2004}. Computational studies include 
overpressurized variable-cell MD~\cite{Nga1992,Nakagiri1982} and transition path sampling by Zahn \& Leoni~\cite{Zahn2004}. 

We describe \ce{NaCl} by the well-known and computationally simple Born-Mayer-Huggins-Fumi-Tosi (BMHFT) potential~\cite{Fumi1964,Tosi1964}, which yields for the equilibrium 
transition pressure at $T=0$ a value of $p_{\text{eq}} = \SI{19.25}{\giga\pascal}$. 
Details of the simulation are provided in the Supplemental Material~\footnote{\label{SM}See 
Supplemental Material at [URL will be inserted by publisher].}.

\begin{figure}[t!]
    \includegraphics[width=7cm]{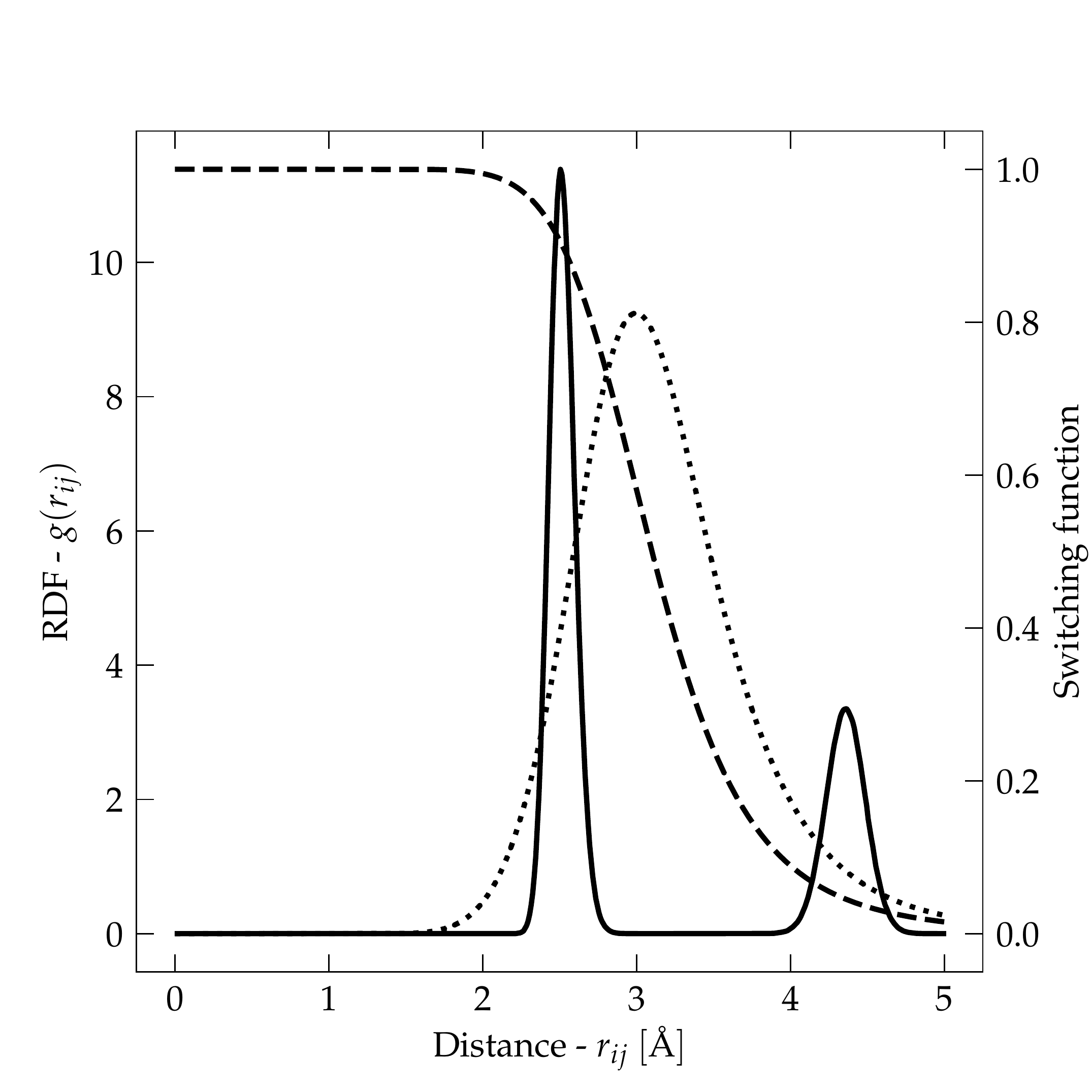}
\caption{The \ce{Na+}-\ce{Cl-} RDF (full) of the $B$1 phase at $p = \SI{20}{\giga\pascal}$ and $T = \SI{300}{\kelvin}$, shown together with the switching function employed (dashed) and the absolute value of its derivative 
(dotted). Note the overlap of the
derivative with the 1\textsuperscript{st} and 2\textsuperscript{nd} coordination spheres. The parameters of the switching function are $d_0 = \SI{1.3}{\angstrom}$ and $r_0 = \SI{2.1}{\angstrom}$.\label{fig:CN_RDF}}
\end{figure}

Structurally, the $B$1-$B$2 transition in \ce{NaCl} is accompanied by the transfer of two ions with opposite charges from the 2\textsuperscript{nd} to
the 1\textsuperscript{st} coordination shell, increasing the CN from 6 in $B$1 to 8 in $B$2. The average CN between
the \ce{Na+} and \ce{Cl-} ions can be calculated by means of a switching function as
\begin{equation}
\mathrm{CN} = \frac{2}{N}\mathlarger{\sum_{\substack{i \in \mathrm{Na}^{+}\\ j \in \mathrm{Cl}^{-}}}}{\left(1+{\left(\frac{r_{ij}-d_0}{r_0}\right)}^6\right)}^{-1}\,\text{,}
\end{equation}
where $r_{ij}$ is the distance between the $i$th cation and the $j$th anion and $N$ is the total number of atoms. The choice of the parameters $d_0$ and $r_0$ 
requires some attention. The switching function should allow to clearly differentiate between the initial state (e.g., $B$1), the transition state, 
and the final state (e.g., $B$2)~\cite{Bussi2020}.  Moreover, its slope should be sufficiently high at the positions of the radial distribution function (RDF) peaks of the $B$1 phase corresponding
to the first and the second coordination sphere in order to drive an easy exchange of ions between the two spheres.
A suitable switching function meeting both requirements is shown in Fig.~\ref{fig:CN_RDF} \footnote{We note that switching functions 
satisfying the second requirement change the numerical value of CN in the respective phases, e.g. in $B$1, 6 $\rightarrow$ 6.9 and in $B$2, 8 $\rightarrow$ 8.1.
For simplicity, we present the true calculated values as such and do not rescale them to neither 6 nor 8.}.

\begin{figure}[t!]
    \includegraphics[width=9cm]{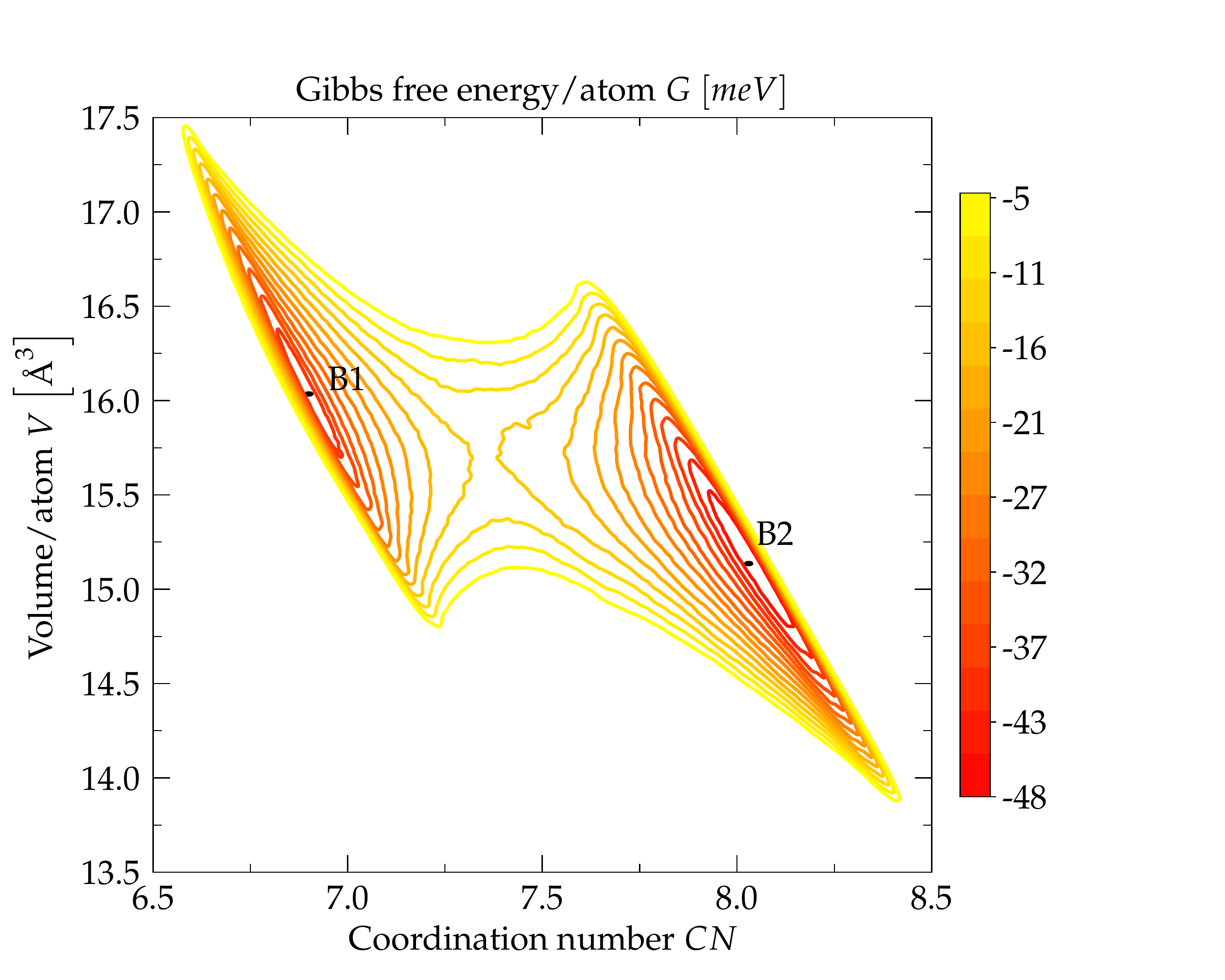}
\caption{(Color online) Reconstructed FES from \SI{100}{\nano\second} MetaD simulation of a 512 atoms system, using CN and $V$ as CVs, at $T = \SI{300}{\kelvin}$ and $p = \SI{20}{\giga\pascal}$. Gaussians of height $\SI{0.41}{\milli\electronvolt\per\mathrm{atom}}$ and width of 0.02 along the CN and $\SI{0.02}{\AA^3}$ 
along the volume CV respectively, were used. The positions of the $B$1 and $B$2 phases are denoted.\label{fig:MetaD_CN_V_GES} }
\end{figure}

For a system in the $B$1 phase with 512 atoms, we performed both MetaD with only CN as well as one with CN \& $V$ as CVs. In both versions, both forward and reverse transitions can 
be seen; see Suplemental Material~\cite{Note1}, Figs. S4-S7. However, the character of the CN evolution in the two cases is different. When 
only the CN is used as CV, even after the first forward and reverse transitions, the system continues to jump between the two phases indicating that the CN does not have full
control over the system. On the other hand, when V is added, the evolution of CN and $V$ after the 1\textsuperscript{st} transitions becomes much more diffusive.
This can be seen in the cross-correlation between CN and $V$; see Supplemental Material~\cite{Note1}, Figs. S8 and S9.
We conclude that CN and $V$ thus represent a good choice of CVs. However, the reconstructed FES in Fig.~\ref{fig:MetaD_CN_V_GES} shows that the
structural phases are represented as rather long and narrow valleys. The soft direction (SD) represents "breathing" of the crystal preserving the structure,
while the perpendicular - hard one (HD) represents a direction of structural change. To improve sampling of such a shaped FES, we introduce a rotation
of CVs with origin at the equilibrium point $[\overline{CN},\overline{V}](p,T)$ of $B$1. Deposited Gaussians thus respect the shape of the valleys, being wide
in the SD and narrow in the HD. We first rescale CN and $V$ with respect to $B$1 and then rotate them by an orthogonal transformation, whose components are  
orthonormal eigenvectors of the covariance matrix of the rescaled coordinates. The covariance matrix  was obtained from a short \SI{200}{\pico\second}
unbiased NPT MD simulation at given pressure $p$ and temperature $T$ in the $B$1 phase. A detailed description of the approach is provided in the Supplemental Material \cite{Note1}.

For illustration, the evolution of the structure of the system across the transition at \SI{40}{\giga\pascal} is shown in the Supplemental Material \cite{Note1}, Fig. S21. 
The intermediate state ($d$) is similar to the $B$33 structure that appears in some theoretically proposed mechanisms (see later). It is seen that the whole system first 
transforms to this transient short-living state which quickly converts to $B$2, pointing to a collective mechanism of the transition. For some alkali-halides,
a two-step transition mechanism through the intermediate bulk $B$16 or $B$33 phases was proposed by Toledáno~\textit{et al.}~\cite{Toldano2003}.

The presented scheme can be readily applied to larger systems allowing to study precursor effects, nucleation and growth and access to information about free energy, size, 
shape, and structure of the critical nucleus. We performed the simulations at  $T = \SI{300}{\kelvin}$ for various system sizes, $N =$ 512, 4\,096, 13\,824, 32\,768 and
64\,000 atoms. Because the size of the critical nucleus at \SI{20}{\giga\pascal}, close to equilibrium, is expected to be very large, we chose to work at pressures of \SI{30}{} and 
\SI{40}{\giga\pascal}. We note that non-classical nucleation theories (see later)~\cite{Moran1996} predict also divergence of the size of the critical nucleus upon approaching the 
point of dynamical instability (for our system we found this at \SI{60}{\giga\pascal}).

\begin{figure}[t!]
    \includegraphics[width=8.6cm]{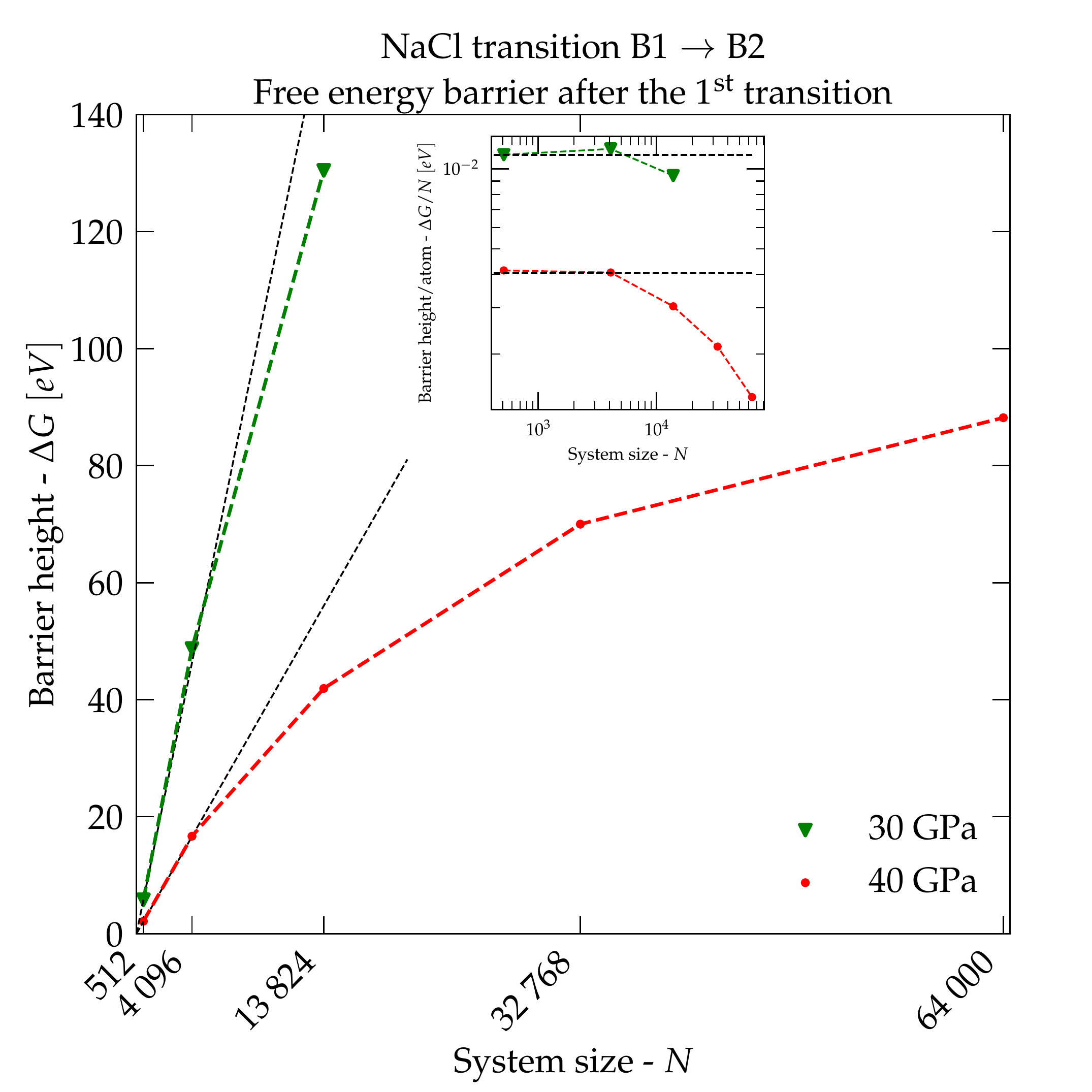}
    \caption{(Color online) Barrier heights (from the $B$1 phase) for various system sizes at $T = \SI{300}{\kelvin}$ \& $p = \SI{30}{}\, \& \SI{40}{\giga\pascal}$. Straight dotted lines represent the values of the barrier from Buerger collective mechanism. The inset shows the barrier height divided by the system size - $\Delta G/N$ vs the system size $N$ on a log-log plot, highlighting the deviations from the linear scaling characteristic of the collective regime. For larger systems, transformation via nucleation and growth proceeds via lower barrier than for the collective mechanism.\label{fig:barrier_vs_system_size}}
\end{figure}

In Fig.~\ref{fig:barrier_vs_system_size} we show the evolution of the transition barrier as a function of system size for two values of pressure~\footnote{The barrier height was determined from Gaussians accumulated in MetaD up to the 1\textsuperscript{st} transition. While it is possible to determine it in more accurate manner, e.g. using the well-tempered metadynamics method~\cite{PhysRevLett.100.020603}, the accuracy of our simple approach is sufficient for the purpose of our study.}. For $p = \SI{30}{\giga\pascal}$ the curve  appears to grow in a nearly linear manner up to $N=13\,824$,
indicating that even at this moderate overpressurization very large system sizes are necessary to properly accommodate the large critical nucleus.
For systems smaller than $4096$ atoms, the barrier per atom agrees well with the estimate based on the static Buerger mechanism 
(see Supplemental Material \cite{Note1}, Fig. S3), showing that the transition proceeds via a collective mechanism. The barrier height in 
the thermodynamic limit must be larger than $10^2\,\SI{}{\electronvolt}$, revealing that homogeneous nucleation in such a regime is physically impossible.
At the higher pressure of \SI{40}{\giga\pascal} the curve appears to eventually converge to a value above \SI{90}{\electronvolt}, still too high for a physical transition. Since 
experimentally the transition at 300 K occurs at $p = \SI{26.6}{\giga\pascal}$~\cite{Li1987}, it must be assisted by extrinsic factors such as lattice 
defects~\cite{Russell1980,Kostorz2001,Fultz2020,Clapp1993,Khaliullin_2011},  dislocations~\cite{Cahn1957,Cook1973,Olson1976_PartI,Olson1976_PartII,Suezawa1980,Li2004,Samanta2014,Levitas2018_dislocation}, grain 
boundaries~\cite{Olson1976_Part_III}, surfaces~\cite{Russell1980,Kostorz2001,Fultz2020}, or non-hydrostatic pressure~\cite{Levitas2020,Levitas2019}. This observation is 
similar to the one found for nucleation of melting~\cite{Samanta2014}, crystallisation of ice~\cite{Sanz2013} and transformation of graphite to diamond~\cite{Khaliullin_2011}. The slow convergence of the barriers can be explained by 
the presence of long-range ($\sim1/r^3$) elastic strain fields~\cite{Nabarro1940,Eshelby1957,Russell1980,Olson1982b,Kostorz2001,Fultz2020}. We note that the elastic energy of the nucleus and
surrounding lattice~\cite{Russell1980,Olson1982b,Kostorz2001,Fultz2020} is taken into account in non-classical nucleation 
theory~\cite{Clapp1973,Clapp1979,Gunin1982,Olson1982b,Gooding1989,Falk1990,Krumhansl1992,Moran1996,Wang1997, Roy1998,Reid1999,Chu2000,Shen2007,Heo2014,Zhang2016} but
is missing in standard static approaches~\cite{Shoji1931,smoluchowski,HydeOKeeffe,Watanabe1977,Pendas1994,Stokes2002,Stokes2004,Sims1998,Toldano2003} which assume a strictly collective 
character of the transformation with no interface between the parent and the new phase. 
It would be fully taken into account in simulation provided the system is sufficiently large.

\begin{figure*}[t!]
	\includegraphics[width=0.99\textwidth,keepaspectratio]{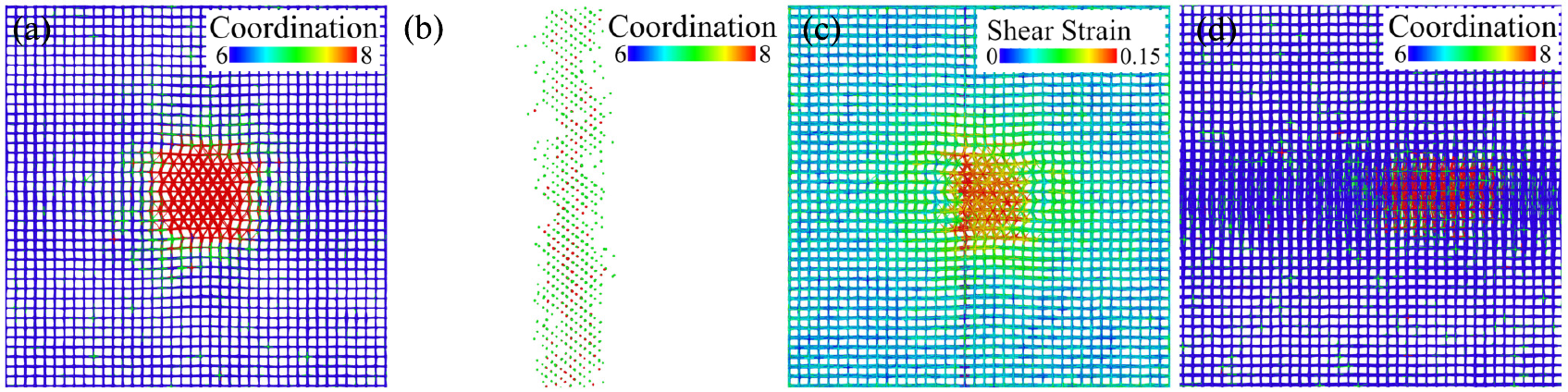}
	\caption{(Color online) Critical nucleus of the $B$2 phase (with the shape of a cylinder)  in the 64000 atoms system at \SI{40}{\giga\pascal}. (a) Critical nucleus (axis of
the cylinder perpendicular to the plane). (b) Perpendicular view of the critical nucleus where only atoms with coordination $ \ge 6.5$ are shown for clarity. (c) Shear around the
critical nucleus in the plane perpendicular to the axis of the cylinder. (d) A localized nucleus with the shape of an ellipsoid (red), 41.7 ps prior to the critical nucleus frame (a). The ellipsoid
grows into the cylinder along the axis in which the strain field extends across the PBC (distortion along horizontal direction). View (d) is perpendicular to both (a) and (b). The pictures were produced using OVITO~\cite{OVITO}. \label{fig:nucleation}}
\end{figure*}

We now focus on the structural aspects of the transition. In Fig.~\ref{fig:nucleation} we see the critical nucleus (determined as the first timestep from which
an unbiased MD proceeds towards the $B$2 basin) in the system of 64\,000 atoms at \SI{40}{\giga\pascal}. Even at this system size, the critical nucleus represents a cylinder 
extending across the periodic boundary conditions (PBC) along one dimension. In all simulations, the nucleus formation starts by creation of strain in a large region of the lattice that extends across the PBC. In this region the primary nucleus is eventually formed, followed by the creation of a secondary nucleus. At all system sizes and pressures presented, the size of the critical nucleus is not small compared to the system size and PBC artefacts are present.
The dependence of shape and size of the critical nucleus on system size and pressure can be found in~the~Supplemental Material \cite{Note1}.

\begin{figure*}[t!]
    \includegraphics[width=0.99\textwidth,keepaspectratio]{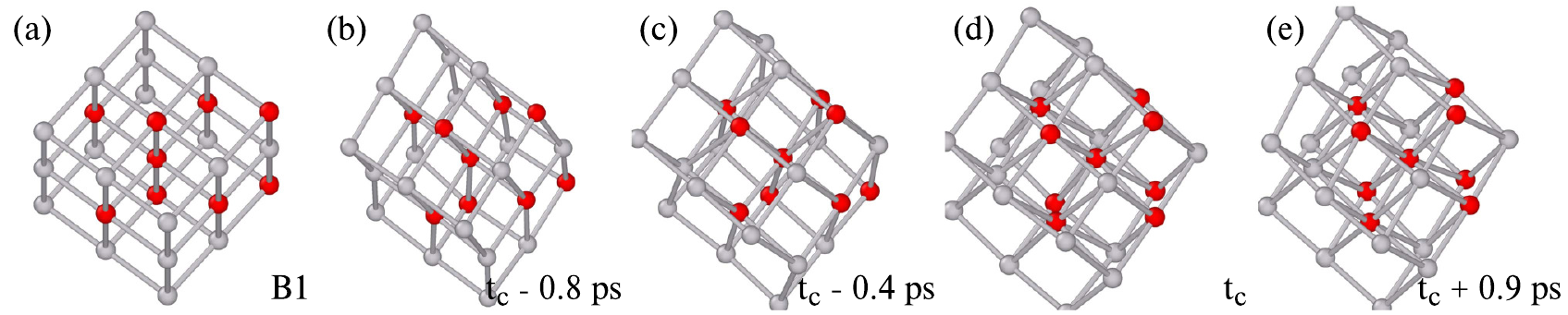}
    \caption{(Color online) WTM-like mechanism (Toledáno) observed 
    during the formation and growth of the critical nucleus in the 64000 atoms system at \SI{40}{\giga\pascal}. The central atom and its 1\textsuperscript{st} and 2\textsuperscript{nd} coordination shell are shown without distinguishing Na and Cl atoms while atoms forming the conventional unit cell of the emerging $B$2 phase are shown in red.
    All atoms are within the critical nucleus and the cutoff for bonds is set to \SI{3}{\angstrom}. The position of the central atom and the point of view are fixed. (a) Atoms in the $B$1 lattice. (b)-(e) Atomic configuration at specified times where (d) corresponds to the critical nucleus. The pictures were produced 
using OVITO~\cite{OVITO}.\label{fig:mechanism}}
\end{figure*}

We further analyzed the detailed transformation mechanism and its dependence on pressure and system size.
For convenience, we provide in the Supplement Material \cite{Note1} a review of previous results found in the literature.
The four idealised collective mechanisms proposed in Ref.~\cite{Stokes2004} can be characterised based on the transformation of the local environment of each atom. $B$2 is formed 
from $B$1 after adding two second neighbours of the opposite type to the first coordination shell. All eight of these second neighbours 
are corners of the conventional \texttt{fcc} cell with a given atom in the centre. If both these additional second neighbours join at the same time and originally form an edge 
of the conventional \texttt{fcc} cell, one finds the WTM mechanism~\cite{Watanabe1977}. On the contrary, if they are located at opposite corners, one finds the Buerger
mechanism~\cite{smoluchowski}. Adding the two atoms independently in two steps instead results in the Toledáno~\cite{Toldano2003} (modified WTM) and Stokes and Hatch 
(modified Buerger) mechanisms, both of which create an intermediate $B$33-like structure~\footnote{The detailed description of the algorithm determining the transformation mechanism 
of the environment of each atom is in the Supplemental Material \cite{Note1}. It is also necessary to distinguish between the mechanism involved 
in the initial creation of the critical nucleus and the one of the subsequent growth.}. This analysis allows the possibility of different parts of 
the system transforming at different times via distinct mechanisms. It was performed for all systems considered in the Supplemental Material \cite{Note1}. 
The WTM and Toledáno mechanisms are facilitated by the local lattice shear strain which amounts to a compression along a $\left<110\right>$ direction. This breaks the cubic symmetry and brings
four out of eight second neighbours closer to the central atom [see Fig. \ref{fig:mechanism}(b)]. It is likely that an application of such uniaxial stress in experiments would
reduce hysteresis and facilitate the observation of the transition closer to the thermodynamic transition pressure.

In our simulations, for systems up to 4096 atoms, the dominant mechanism is related to the creation of the intermediate (bulk) $B$33 structure that subsequently transforms to $B$2  
via the Stokes and Hatch mechanism. For a system size of 4096 atoms, only parts of the system locally transform through the Stokes and Hatch mechanism,
see Supplemental Material \cite{Note1}, Fig. S28(e). Finally, for systems larger than 4096 atoms, all atoms within the critical 
nucleus transform via the Toledáno mechanism. This involves the displacement of planes, as can be seen in Fig.~\ref{fig:mechanism}(c). 
In larger systems the Stokes and Hatch mechanism would cost too much energy and therefore nucleation via a zig-zag pattern (WTM or Toledáno mechanism),
which causes less strain, appears to be preferable. Nuclei are surrounded by 7-coordinated atoms, but this layer does not resemble $B$33-like structures in large systems. 

Our approach is likely to work for a broader class of pressure-induced structural transitions and uncover their microscopic mechanisms 
in the regime of nucleation and growth, including calculation of free-energy barriers. It might represent a bridge between atomistic modelling of structural phase transitions
and effective phase-field 
theories~\cite{Clapp1973,Clapp1979,Gunin1982,Olson1982b,Gooding1989,Falk1990,Krumhansl1992,Moran1996,Wang1997,Roy1998,Reid1999,Chu2000,Shen2007,Heo2014,Zhang2016,Levitas2002,Levitas2002_2,Levitas2003,Zhang2007,She2013,Levitas2014,Levitas2016,Levitas2017,Levitas2018,Levitas2018_2}. 
The use of simple and physically naturally motivated CVs allows a MetaD simulation without prior knowledge of the transition and the final states. This would 
enhance the predictive value, in particular in cases in which the final state might be either a metastable one or one which is stabilized by entropy and therefore falls beyond the reach of 
standard $T=0$ structural prediction methods. It is essential to access long time ($\sim$ 10 ns) as well as length scales (more than $10^5$ atoms) which 
should be feasible by using machine learning-based potentials~\cite{Noe2020,Behler2007,Behler2014,DellagoBehler,Behler2021,Car,Car2,Car2019,Csanyi,Csayni2,Tchatenko,Shaidu2021}. 
The approach can be generalized to study the role of non-hydrostatic pressure (similarly to Ref.~\cite{Donadio2008}),
important  in diamond-anvil-cell experiments. Our results point to the need for studying structural transitions in non-idealized environments closer to realistic conditions,
including the presence of structural defects, such as dislocations.

We thank G. Bussi, A. Laio, S. Leoni and C. Molteni for the careful reading of the manuscript and useful comments. RM acknowledges stimulating discussions with M. Parrinello, E. Tosatti, A. Laio and
S. Scandolo. MB acknowledges all answers to his questions on PLUMED, LAMMPS and DLPOLY users mailing lists during the course of this work,
in particular to G. Bussi and A. M. Elena, as well as funding provided by FMFI UK to attend the NCCR-MARVEL School on VES in Feb 2017.
This work was supported by the Slovak Research and Development Agency under Contracts APVV-15-0496 and APVV-19-0371, by VEGA project 1/0640/20
and by Comenius University under grant for young researchers - UK/436/2021.

%

\end{document}